\pdfoutput=1

\documentclass[aps,prl,twocolumn,showpacs,preprintnumbers,fleqn,superscriptaddress,floatfix]{revtex4-1}

\usepackage{graphicx}
\PassOptionsToPackage{colorlinks=true,linkcolor=magenta,citecolor=green,urlcolor=blue}{hyperref}
\usepackage{hyperref}

\begin{document}

\preprint{%
published in Phys. Rev. Lett. \textbf{112}, 156101 (2014)
DOI: \href{http://dx.doi.org/10.1103/PhysRevLett.112.156101}{10.1103/PhysRevLett.112.156101}
}

\title{Role of Physisorption States in Molecular Scattering:\\
A Semilocal Density-Functional Theory Study on O$_{2}$/Ag(111)}

\author{I. Goikoetxea}
\altaffiliation{Humboldt Universit\"at zu Berlin, Institut f\"ur Chemie, Unter den Linden 6, D-10009 Berlin, Germany}
\affiliation{Centro de F\'{\i}sica de Materiales CFM/MPC (CSIC-UPV/EHU),
Paseo Manuel de Lardizabal 5, E-20018 San Sebasti\'an, Spain}

\author{J. Meyer}
\altaffiliation{Leiden Institute of Chemistry, Gorlaeus Laboratories, Leiden University, P.O. Box 9502, 2300 RA Leiden, The Netherlands}
\affiliation{Chair for Theoretical Chemistry and Catalysis Research Center,\\ Technische Universit{\"a}t M{\"u}nchen, Lichtenbergstr. 4, D-85747 Garching, Germany}

\author{J.I.Juaristi}
\affiliation{Centro de F\'{\i}sica de Materiales CFM/MPC (CSIC-UPV/EHU),
Paseo Manuel de Lardizabal 5, E-20018 San Sebasti\'an, Spain}
\affiliation{Departamento de F\'{\i}sica de Materiales, Facultad de
Qu\'{\i}micas, UPV/EHU, Apartado 1072, E-20080 San Sebasti\'an, Spain}
\affiliation{Donostia International Physics Center DIPC, Paseo Manuel de
Lardizabal 4, E-20018 San Sebasti\'an, Spain}

\author{M. Alducin}
\affiliation{Centro de F\'{\i}sica de Materiales CFM/MPC (CSIC-UPV/EHU),
Paseo Manuel de Lardizabal 5, E-20018 San Sebasti\'an, Spain}
\affiliation{Donostia International Physics Center DIPC, Paseo Manuel de
Lardizabal 4, E-20018 San Sebasti\'an, Spain}

\author{K. Reuter}
\affiliation{Chair for Theoretical Chemistry and Catalysis Research Center,\\ Technische Universit{\"a}t M{\"u}nchen, Lichtenbergstr. 4, D-85747 Garching, Germany}

\date{\today}

\begin{abstract}
We simulate the scattering of O$_2$ from Ag(111) with classical dynamics
simulations performed on a six-dimensional potential energy surface calculated
within semi-local density-functional theory (DFT). The enigmatic experimental trends
that originally required the conjecture of two types of repulsive walls, arising
from a physisorption and chemisorption part of the interaction potential, are
fully reproduced. Given the inadequate description of the physisorption properties
in semi-local DFT, our work casts severe doubts on the prevalent notion to use
molecular scattering data as indirect evidence for the existence of such states.
\end{abstract}

\pacs{34.35.+a,68.49.Df,82.20.Kh}

\maketitle

Molecular scattering experiments at single-crystal surfaces have a long
tradition in the analysis of molecule-surface interaction and energy loss
mechanisms, which are both key aspects in the quest to understand gas surface
dynamics~\cite{toennies74,kleyn03,vattuone10}. A prevalent notion in the
analysis of this kind of scattering data is that it is critically influenced
by the potential energy surface (PES) topology at large distances from the
surface. As physisorption states are located a such large distances, there
is a general perception that scattering experiments are rather sensitive to
them, or vice versa that scattering experiments provide detailed information
on such kind of states.

In this understanding, it is puzzling to realize that several
recent first-principles studies modeling the scattering
dynamics at metal surfaces have reached quite satisfactory,
if not quantitative descriptions of corresponding experimental
data~\cite{nieto06,diaz06,geethalakshmi11,batista11,nattino12}. In these
studies the high-dimensional PES underlying the dynamical simulations is
generally obtained from density-functional theory (DFT) with semi-local
exchange-correlation (xc) functionals. By construction, such short-ranged
functionals cannot provide an adequate description of van der Waals
interactions~\cite{kristyan94,hult96,kohn98,rydberg03}, and the corresponding
PES will not feature any weakly-bound physisorption states far away from
the surface.

In order to further elucidate this point we focus here on
O$_2$/Ag(111), which is not only a prototypical system allegedly
featuring a long discussed (but still rather elusive) physisorption
state~\cite{campbell85,schmeisser85,spruit89,besenbacher93,lacombe96}, but
also a system where detailed scattering data have been centrally interpreted
in terms of a repulsive wall separating this physisorption from a stronger
bound chemisorption state. In contrast, the PES from semi-local DFT that
has already been used to successfully describe the molecular sticking data
at this surface~\cite{goikoetxea12} does not feature a physisorption well,
and as such this system represents an intriguing test case to pinpoint the
role of physisorption states for molecular scattering.

The scattering of O$_2$ from Ag(111) is simulated following the
divide-and-conquer approach we already used in the study of the O$_2$
dissociative adsorption on this surface~\cite{goikoetxea12}. This approach
requires as a first step the calculation of an accurate multi-dimensional PES,
which is subsequently used for evaluating the gas-surface dynamics. Here,
the ground-state PES along the six O$_2$ molecular degrees of freedom
at a rigid Ag(111) surface is first extensively mapped with DFT
calculations performed with the semi-local xc functional due to Perdew,
Burke and Ernzerhof (PBE)~\cite{perdew96}. Infinite molecule-surface
separation defines the zero energy reference. A neural networks-based
technique~\cite{lorenz04,lorenz06} that properly exploits the symmetry of
the studied system~\cite{behler07,goikoetxea12} then yields a continuous
six-dimensional (6D) PES. On this PES representation millions of classical
molecular dynamics trajectories are finally integrated at low computational
cost.

The scattering of O$_2$ from Ag(111) was extensively studied by Raukema {\em
et al.} using a supersonic molecular beam apparatus that permitted incidence
energies $0.4 < E_{\rm i} < 1.8$\,eV at defined incidence angles $(\Theta_{\rm
i}, \Phi_{\rm i})$~\cite{raukema95}. Compared with their results for N$_2$
scattering and largely independent of the azimuthal angle $\Phi_{\rm i}$,
the in-plane angular distributions $\Theta_{\rm f}$ of the scattered O$_2$
molecules exhibit a peculiar variation with $E_{\rm i}$ and $\Theta_{\rm
i}$: At $\Theta_{\rm i}=40^{\circ}$, cf. Fig. 1 for a definition of the
scattering geometry, the distribution is narrowly peaked around specular
reflection for low incidence energies, $E_{\rm i}\simeq 0.4$~eV. At $E_{\rm
i}$ increasing above 1.0~eV the distribution instead rapidly broadens towards
the surface normal, concomitant with a decreasing intensity at specular
reflection. This $E_{\rm i}$-dependence is intriguingly smoothly reversed,
when the polar incidence angle varies from $40^{\circ}$ to $70^{\circ}$. At
the largest angle ($\Theta_{\rm i} =70^{\circ}$) symmetric and narrow angular
distributions around specular are obtained for all incidence energies. However,
in contrast to the $\Theta_{\rm i} =40^{\circ}$ case, it is at the largest
incident energy, when the distribution is the narrowest and the height of
the distribution peak the largest.

In ref.~\cite{raukema95} these results were rationalized in terms of
scattering from a dual repulsive wall arising from the physisorption and
chemisorption part of the interaction potential. In this picture, low
normal-energy molecules are scattered at large distances from the surface
in the repulsive walls connected with the shallow physisorption state,
where the PES is still rather uniform in its lateral degrees of freedom. In
contrast, high normal-energy molecules overcome these first barriers and
are reflected closer to the surface on the repulsive walls related with
the chemisorption state where the PES is more corrugated. The broadening of
the angular distributions for the more normal incidence $\Theta_{\rm i} =
40-60^{\circ}$ is thus explained by an increase in lateral PES corrugation
experienced by the more energetic O$_2$ molecules. In contrast, at the
most grazing $\Theta_{\rm i} = 70^{\circ}$ all molecules would already be
reflected by the walls of the physisorption state. In this case, the varying
width of the angular distributions was associated to thermal broadening,
i.e. the influence of the surface atom vibrations.

Quite common for the interpretation of scattering data, this picture assumes
the existence of two separate adsorption wells, a shallow physisorption state
and a deeper chemisorption state. This is at variance with the previous
analysis of our calculated 6D DFT-PES~\cite{goikoetxea12}, whose high
interpolation quality is guaranteed by the small ($< 26$ meV) root mean square error (RMSE) in
dynamically important regions. There is indeed a molecular chemisorption
well, and a steep repulsive wall with energies of more than 1\,eV required
to bring the molecule closer than 2~{\AA} to the surface. The chemisorption
well is only -40~meV deep (though -70~meV if surface relaxation is allowed),
which already casts severe doubts on its actual role for the scattering
data. Moreover, in the DFT-PES, calculated using a semi-local xc functional,
physisorption wells are entirely absent. In this respect, multi-dimensional
dynamics simulations on this PES are ideally suited to elucidate which PES
topological features really rule the peculiar O$_2$ scattering off Ag(111).

\begin{figure}
\suppressfloats
\centering
\includegraphics[clip,width=0.45\textwidth]{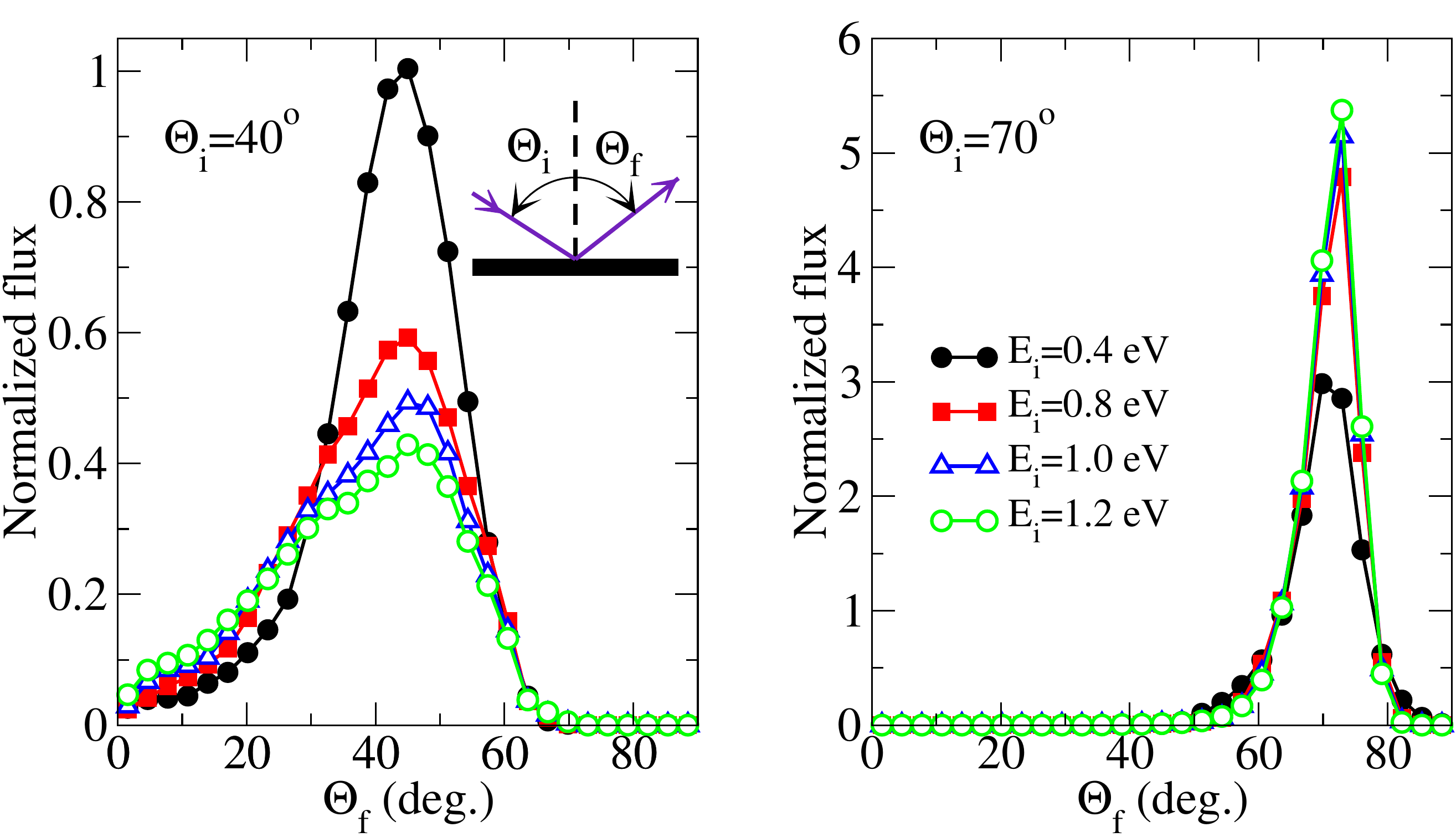}
\caption{(Color online) Calculated angular distributions for in-plane O$_2$
scattering off Ag(111) for incidence angles $\Theta_{\rm i}=40^{\circ}$ (left
panel) and $\Theta_{\rm i}=70^{\circ}$ (right panel) and various $E_{\rm i}$. All
fluxes are normalized to the peak height obtained for $\Theta_{\rm i}=40^{\circ}$
and $E_{\rm i}$=0.4~eV.}
\label{fig:scattering}
\end{figure}

\begin{figure}
\suppressfloats
\centering
\includegraphics[clip,width=0.45\textwidth]{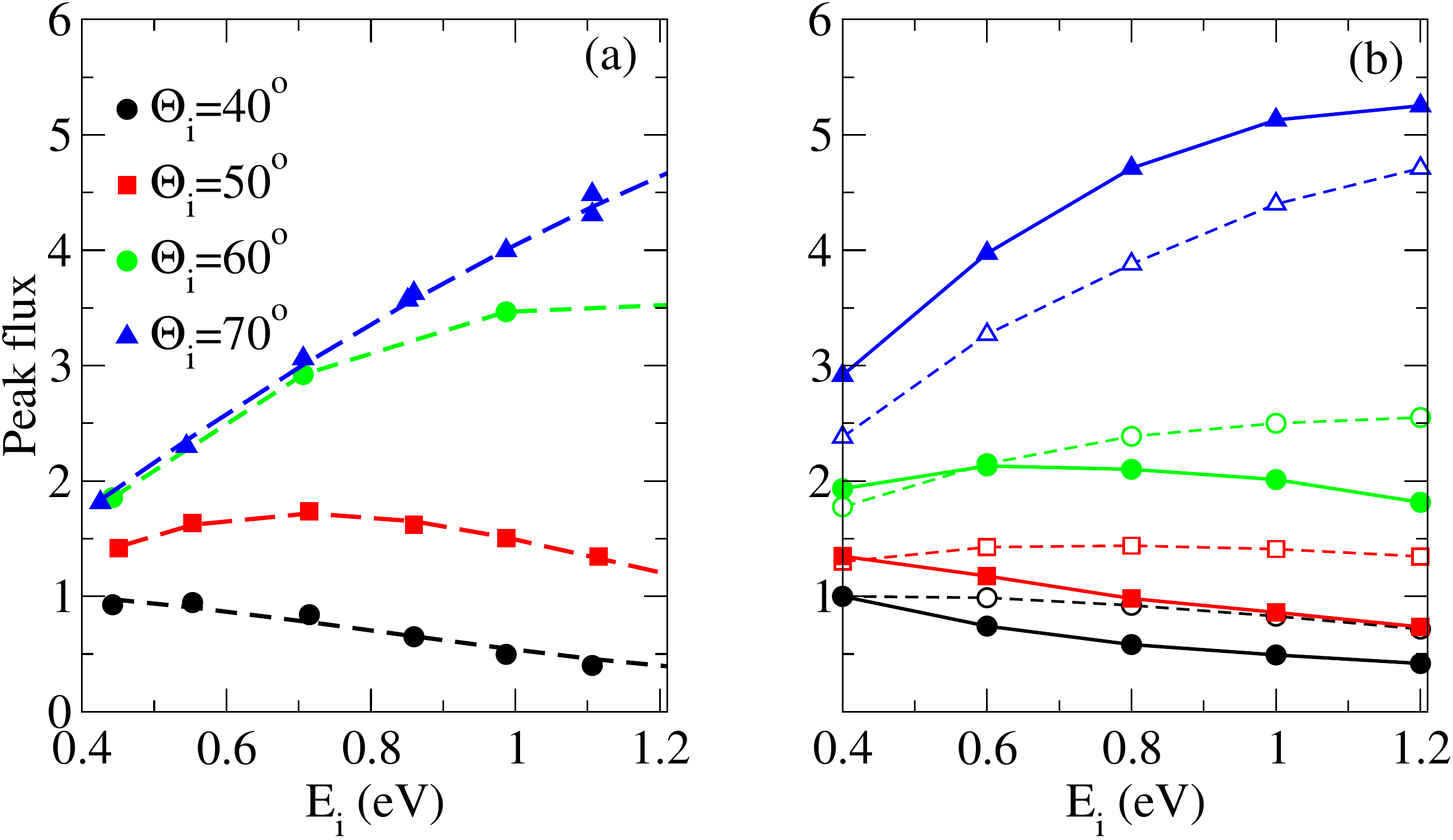}
\caption{(Color online) Peak height of the distributions of
Fig.~\ref{fig:scattering} as a function of incidence energy $E_{\rm i}$. The
experimental data reproduced from Ref.~\cite{raukema95} (left panel) can be directly
compared with the results of our adiabatic (full lines) and GLO (dashed lines)
simulations (right panel). Lines through the data points are guides to the eye.}
\label{fig:peaks}
\end{figure}

With this objective we perform classical trajectory calculations for defined
$E_{\rm i}$ (neglecting the initial zero-point energy) and $\Theta_{\rm
i}$. For each ($\Theta_{\rm i}$,$E_{\rm i}$) condition $10^6$ trajectories
sample over all possible initial O$_2$ angular orientations and lateral
positions over the surface unit cell. Consistent with experiment, we do
not observe any specific dependence on the azimuthal incidence angle. At an
unspecified crystal orientation in the measurements, the trajectory ensemble
correspondingly also averages over this degree of freedom. The in-plane
scattering angle distributions $\Theta_{\rm f}$ obtained from these simulations
are shown in Fig.~\ref{fig:scattering} for the two extreme incidence angles
used in experiment, namely $\Theta_{\rm i}=40^{\circ}$ and $\Theta_{\rm
i}=70^{\circ}$. Following ref.~\cite{raukema95}, all curves are normalized to
the maximum value of the distribution we obtain for $\Theta_{\rm i}=40^{\circ}$
and $E_{\rm i}$=0.4~eV. For each incidence condition ($\Theta_{\rm i},E_{\rm
i}$), the in-plane distribution is calculated by selecting the molecules
that are reflected within a fixed solid angle. This means that for each
$\Theta_{\rm f}$ the acceptance azimuthal angle is given by $|\Phi_{\rm
f}-\Phi_{\rm i}| < \Delta\Phi^0/\sin\Theta_{\rm f}$, where $\Delta\Phi^0$
is the acceptance azimuthal angle at $\Theta_{\rm f}=90^{\circ}$. The value
$\Delta\Phi^0=5^{\circ}$ that has been used in calculating the distributions
assures reliable statistics even at extreme observation conditions where
reflection events are less probable.

In full agreement with experiment we observe that the decrease in the
distribution maximum is accompanied by an increase of the curve width.
In fact, despite the absence of physisorption states in the DFT-PES our
simulations reproduce rather well all general experimental trends described
above. Whereas for $\Theta_{\rm i}=40^{\circ}$ the distributions broaden
towards the surface normal and the height of the specular peak monotonically
decreases as $E_{\rm i}$ increases, for $\Theta_{\rm i}=70^{\circ}$
the broadening is rather symmetric and the peak behavior is just the
opposite. Note that the peak decrease is not only a consequence of the
broadening in the in-plane polar distribution, but also related to an increase
in out-of-plane scattering events. The change in the peak height is explicitly
compared to the experimental data in Fig.~\ref{fig:peaks}. In both cases
(Figs.~\ref{fig:peaks}(a) and (b)) the decrease of the peak height with
increasing $E_{\rm i}$ is gradually turned into an increase as the incidence
angle is varied from 40$^{\circ}$ to 70$^{\circ}$.

\begin{figure}
\centering
\includegraphics[clip,width=0.45\textwidth]{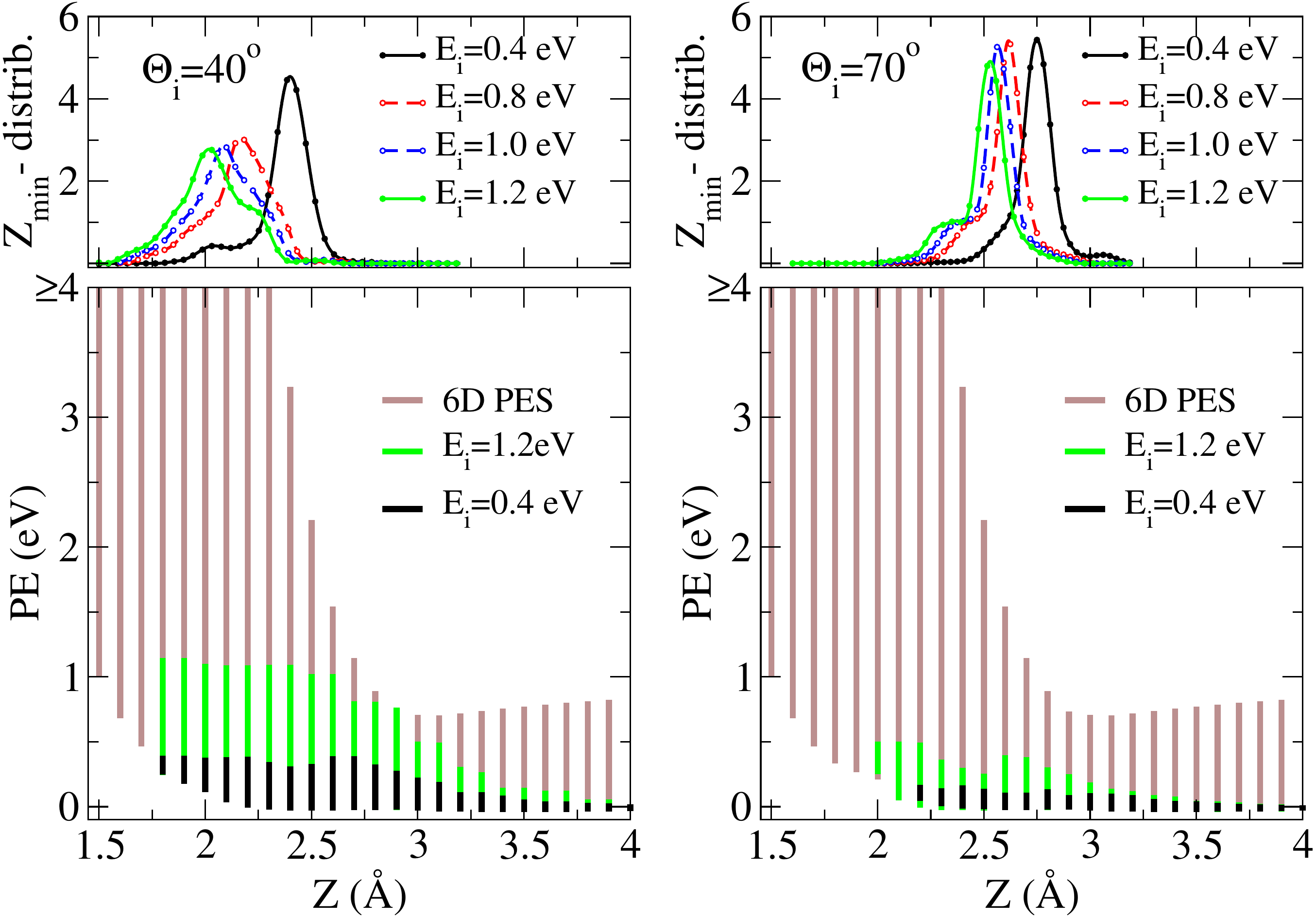}
\caption{(Color online) Upper panels: Distributions of the molecule's
closest approach distance $Z_{\rm min}$ to the surface for the different
incidence conditions. The distribution curve for each $\Theta_{\rm i}$
and $E_{\rm i}$ is normalized to the corresponding total number of
in-plane scattered molecules. Lower panels: Range of PE values exhibited
by the 6D PES $E(X,Y,Z,d,\theta,\phi)$ as a function of distance $Z$ to
surface (maroon vertical lines) as compared to the PE range that is
actually probed by the O$_2$ molecules along their trajectories for
$E_{\rm i}= 0.4$\,eV (black vertical lines) and $E_{\rm i}=1.2$\,eV
(green vertical lines), see text. Left panels correspond to $\Theta_{\rm i}=40^{\circ}$, right panels to $\Theta_{\rm i}=70^{\circ}$. Incidence angles in between (not shown) exhibit a smooth transition between these two limits.}
\label{fig:pesrange}
\end{figure}

The lower panels of Fig.~\ref{fig:pesrange} provide first insight into
the complex and variable energy landscape that the molecules can encounter
while approaching the Ag(111) surface. In the conceptual one-dimensional
picture underlying the experimental data analysis~\cite{raukema95}, the
potential energy (PE) as a function of the distance $Z$ from the surface,
$E = E(Z)$, exhibits two wells separated by a barrier and a repulsive
wall at distances closer than the chemisorption well. Depending on
the O$_2$ orientation ($\theta$, $\phi$) and lateral position $(X,Y)$
over the surface unit cell, the true 6D PES $E(X,Y,Z,d,\theta,\phi)$ as
a function of $Z$ exhibits a range of values stretching over several eV,
even if restricting the O$_2$ internuclear distance $d$ to values between
the one in gas-phase (1.24\,{\AA}) and that at the chemisorption state
(1.28\,{\AA}). At this complexity, it is enlightening to analyze which of
these PE values are actually encountered by the incoming O$_2$ molecules
for the varying incidence conditions. As shown in Fig. \ref{fig:pesrange}
the incidence energy can effectively be used to reach regions of PEs up to
almost $E_i$ for $\Theta_{\rm i}=40^{\circ}$, whereas for the more grazing
$\Theta_{\rm i}=70^{\circ}$ such regions are inaccessible. Most importantly,
the PE regions visited in the latter case always stay below the normal
energy component of $E_{\rm i}$, whereas for $\Theta_{\rm i}=40^{\circ}$
this is only the case for the lowest $E_{\rm i} = 0.4$\,eV. This analysis
thus supports the effective `dual-repulsive' wall picture of Raukema {\em
et al.}: At $\Theta_{\rm i}=70^{\circ}$ (and $\Theta_{\rm i}=40^{\circ},
E_{\rm i} = 0.4$\,eV) the O$_2$ molecules are dominantly scattered at energy
barriers in the entrance channel, i.e. at distances far from the surface
where at low lateral PES corrugation the dynamics is still governed by
the normal energy ~\cite{gadzuk85,riviere06,goikoetxea12b}. At the other
incidence conditions, corresponding to normal energies exceeding 0.25\,eV,
the molecules get closer to the surface and are scattered at more corrugated
PES parts where the normal-energy condition is no longer obeyed.

This view of the scattering at two types of repulsive regions is further
confirmed by the analysis of the distance of closest approach $Z_{\rm min}$
of the in-plane reflected molecules. The results plotted in the upper
panels of Fig.~\ref{fig:pesrange} clearly demonstrate that there are two
distinct regions, where the molecules are preferentially reflected. At
incidence conditions corresponding to low normal energies the $Z_{\rm
min}$--distributions are characterized by a quite narrow peak at distances
above 2.3~{\AA} and a small tail extending to closer distances from the
surface. In these cases, the dominant fraction of the molecules is thus
reflected above 2.3~{\AA} from the surface. In contrast, for incidence
conditions corresponding to normal energies above 0.25\,eV, the distributions
are visibly shifted to smaller $Z_{\rm min}$ values and exhibit a very broad
shape. Such broadening is clearly related to an increase in the corrugation
of the PES close to the surface.

With this understanding, we can thus fully rationalize the inversion of the
peak height variation with incidence energy for the two incidence angles. At
$\Theta_{\rm i}=40^{\circ}$ and $E_{\rm i} >0.4$\,eV the majority of the
molecules are reflected below 2.3~{\AA}, where the lateral PES corrugation
is already important. The observed decrease in the height of the specular
reflection peak at this incidence angle with increasing $E_i$ is thus a
consequence of the larger PES corrugation probed by the molecules getting
closer to the surface. In contrast, at $\Theta_{i}=70^{\circ}$ the normal
energy for all $E_{\rm i}$ is so small, that most of the molecules are
reflected far from the surface without probing the corrugated part of the
PES. Intriguingly, here our simulations at a rigid Ag(111) surface fully
reproduce the decrease of the height of the specular peak with decreasing
$E_{\rm i}$ that was originally ascribed to thermal broadening. To further
scrutinize that surface vibrations are indeed not key to this effect, we
carried out additional simulations using the generalized Langevin oscillator
(GLO) model to mimic surface mobility~\cite{adelman79,tully80}. The simulations
are performed following the implementation of ref.~\cite{busnengo05};
technical details on how the GLO is applied to the Ag(111) surface can be
found in ref.~\cite{martin12}. The GLO results, shown as dashed lines in
Fig.~\ref{fig:peaks}(b), are indeed very similar to those obtained within the
adiabatic frozen surface approximation. This confirms that the peak height
decrease arises mostly from the small PES variations of about 20--30~meV at
distances $Z > 3$~\AA, which become increasingly irrelevant as the normal
incidence energy of the molecules increases. Note that for an incidence
angle of $60^{\circ}$ there is a qualitative difference between our adiabatic
calculations and the GLO results, which reproduce better the experimentally
observed trend. This shows that, though it is not the dominant mechanism,
some effect due to surface vibrations cannot be disregarded.

\begin{figure}
\centering
\includegraphics[clip,width=0.45\textwidth]{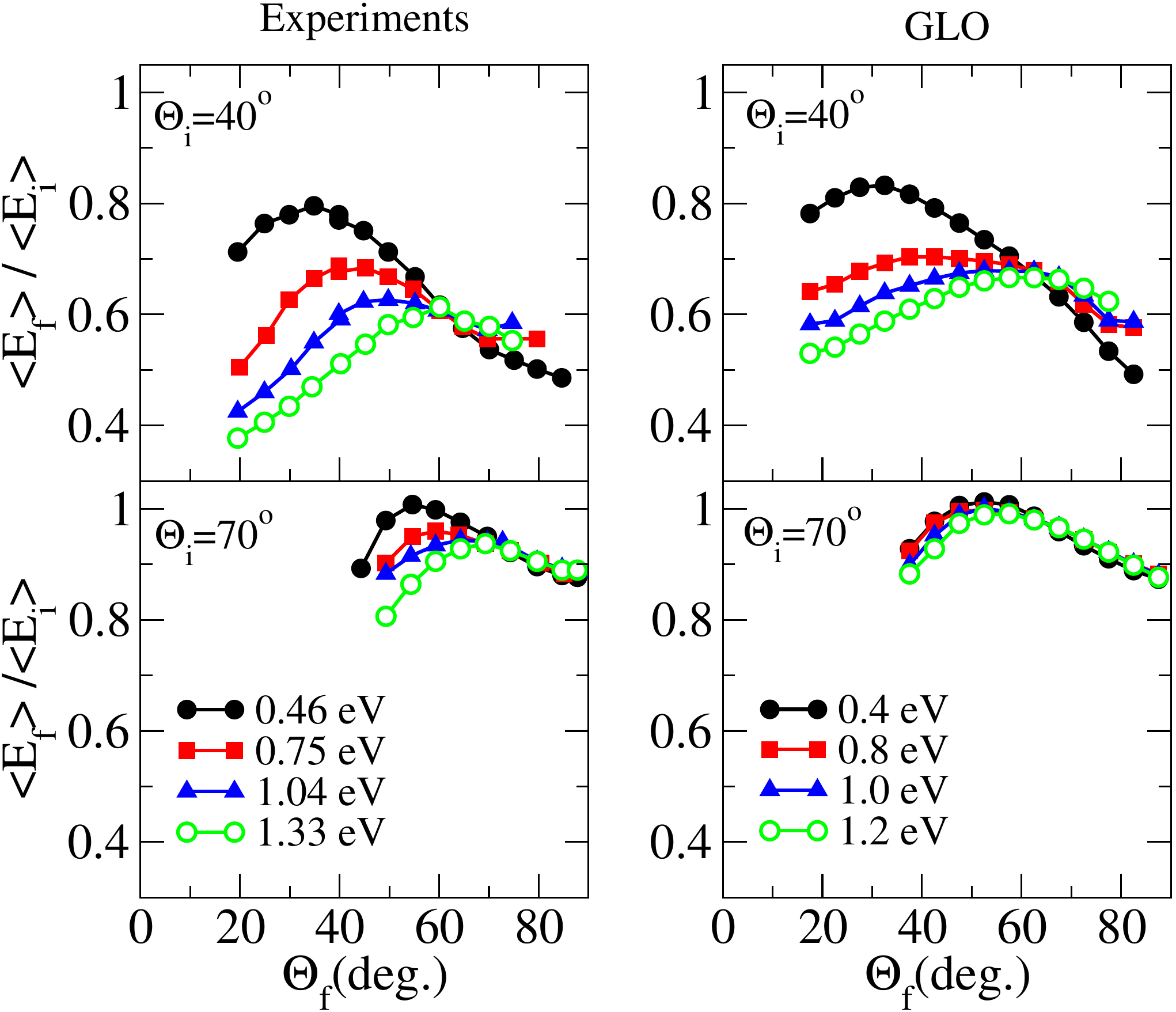}
\caption{(Color online) Angularly resolved final-to-initial energy distributions
of O$_2$ scattered off Ag(111) under different incidence conditions ($\Theta_{\rm
i}$,$E_{\rm i}$). The experimental data by Raukema {\it et al.}~\cite{raukema95}
(left panels) are compared to the results from the GLO calculations (right panels),
see text.}
\label{fig:eloss}
\end{figure}

In addition to the scattering angular distributions, Raukema {\it et al.} measured
the energy loss distribution of the scattered molecules~\cite{raukema95}. These
additional results were also interpreted as a further validation of the
dual repulsive wall picture. The experimental energy loss distributions are
reproduced in Fig.~\ref{fig:eloss}. This figure shows the angularly resolved
final-to-initial translational energy ratio distributions of O$_2$ scattered from
Ag(111) at different incidence conditions. Within this kind of representation,
values of $\langle E_{\rm f}\rangle /\langle E_{\rm i}\rangle < 1$ show that the
molecule has lost translational energy after reflection (the larger the energy
loss, the smaller the energy ratio). For $\Theta_{\rm i}=70^{\circ}$, the energy
loss distributions measured at different incidence energies reasonably follow the
behavior expected for scattering from a flat PES, for which parallel momentum is
conserved. This is precisely the behavior usually identified with physisorption-like
interactions. However, as $\Theta_{\rm i}$ decreases, the final-to-initial energy
curves are no longer subject to parallel momentum conservation. At the smallest
incidence angle ($\Theta_{\rm i}=40^{\circ}$), the O$_2$ scattering is instead
characterized by high energy losses. In fact, under large normal incidence energy
conditions, the measured energy losses for outgoing angles $\Theta_{\rm f}\leq
60^{\circ}$ can exceed those predicted within the binary-collision model, which
is the manifestation of a highly corrugated PES. This result was interpreted as
an indication of the molecules being reflected in a region close to the surface.

The final-to-initial energy distributions calculated with the GLO model are shown
in the right panels of Fig.~\ref{fig:eloss}. In all the cases, the qualitative
agreement between the results of our simulations and the experimental data is
remarkable. In spite of its simplicity, the GLO model reproduces the shape and
trends of the $\langle E_{\rm f}\rangle /\langle E_{\rm i}\rangle$ distributions,
as well as the dependence on $E_{\rm i}$ and $\Theta_{\rm i}$. We state that this
is a further validation of the quality of our PES and its capability to capture
the main characteristics of the scattering dynamics.

In summary, we have studied the scattering properties of O$_2$ molecules incident
on the Ag(111) surface. Simulating the detailed scattering properties at various
incidence angles and energies we can fully reproduce the enigmatic trends that
originally required the conjecture of two types of repulsive walls in the
PES, arising from a physisorption and chemisorption part of the interaction
potential. In fact, a detailed analysis of the simulated trajectories nicely
confirms this concept of two repulsive walls. However, this is in the absence of
any physisorption state in our DFT-PES. The solution to this puzzle comes with
the realization that scattering is not sensitive to the physisorption well itself,
but instead to the repulsive wall ''behind" a well. These walls typically emerge
out of rehybridization phenomena (the molecular states need to orthorgonalize to
the electronic states of the surface) and semi-local DFT can describe this aspect
rather reliably. Furthermore, it is only in an effective one-dimensional picture
that a wall necessarily separates two distinct wells. In higher dimensions this is
no longer true, and the existence of different repulsive regions in the PES must
not necessarily imply the existence of a separate physisorption state. As such,
our work casts severe doubts on the prevalent notion to use molecular scattering
data as indirect evidence for the existence of such states. As with many other
techniques, prominently temperature-programmed desorption, such weakly-bound
states far away from the surface remain hard to grasp quantitatively and their
real relevance for gas-surface dynamics remains uncertain.

\begin{acknowledgments}
The work of I.G. has been supported by the Spanish Research Council (CFM-CSIC,
Grant Nos. JAE-Pre\_08\_0045 and 2010ESTCSIC-02167). Funding by the Deutsche
Forschungsgemeinschaft is gratefully acknowledged. M.A. and J.I.J. acknowledge
financial support by the Basque Departamento de Educaci\'on, Universidades
e Investigaci\'on, the University of the Basque Country UPV/EHU (Grant
No. IT-756-13) and the Spanish Ministerio de Ciencia e Innovaci\'on (Grant
No. FIS2010-19609-C02-02).
\end{acknowledgments}

\bibliographystyle{apsrev4-1}
\bibliography{o2ag111scatt}

\end{document}